\documentclass{sf2a-conf2017}
\usepackage{graphicx}
\usepackage{hyperref}
\usepackage[]{natbib}  
\usepackage{epstopdf}

\def\BibTeX{{\rm B\kern-.05em{\sc i\kern-.025em b}\kern-.08em
    T\kern-.1667em\lower.7ex\hbox{E}\kern-.125emX}}
\bibpunct{(}{)}{;}{a}{}{,}  

\begin{document}

\TitreGlobal{SF2A 2017}


\title{Radio morphing - towards a full parametrisation of the radio signal from air showers}
\runningtitle{Radio morphing}

\author{Anne Zilles}\address{Sorbonne Universit\'{e}s, UPMC Univ.  Paris 6 et CNRS, UMR 7095, Institut d'Astrophysique de Paris, 98 bis bd Arago, 75014 Paris, France}
\author{Didier Charrier}\address{SUBATECH, IN2P3-CNRS, Universit\'{e} de Nantes, Ecole des Mines de Nantes, Nantes, France}
\author{Kumiko Kotera}\address{Sorbonne Universit\'{e}s, UPMC Univ.  Paris 6 et CNRS, UMR 7095, Institut d'Astrophysique de Paris, 98 bis bd Arago, 75014 Paris, France et  Laboratoire AIM-Paris-Saclay, CEA/DSM/IRFU, CNRS, Universit\'{e} Paris Diderot, F-91191 Gif-sur-Yvette, France} 
\author{Sandra Le Coz}\address{National Astronomical Observatories, Chinese Academy of Sciences, Beijing 100012, China} 
\author{Olivier Martineau-Huynh}\address{LPNHE, CNRS-IN2P3 et Universit\'{e}s Paris VI \& VII, 4 place Jussieu, 75252 Paris, France}
\author{Clementina Medina}\address{APC Laboratory, Universit\'{e} Paris Diderot /  Instituto Argentino de Radioastronom\'{i}a - CONICET / CCT-La Plata, Camino Gral. Belgrano Km 40, Berazategui, Buenos Aires, Argentina} 
\author{Valentin Niess}\address{Clermont Universit\'{e}, Universit\'{e} Blaise Pascal, CNRS/IN2P3, Laboratoire de Physique Corpusculaire, BP. 10448, 63000 Clermond-Ferrand, France}
\author{Matias Tueros}\address{Instituto de F\'{i}sica La Plata - CONICET/CCT- La Plata. Calle 49 esq 115. La Plata, Buenos Aires, Argentina}
\author{Krijn de Vries}\address{Vrije Universiteit Brussel, Physics Department, Pleinlaan 2, 1050 Brussels, Belgium}

\setcounter{page}{237}


\maketitle


\begin{abstract}
Over the last decades, radio detection of air showers has been established as a detection technique for ultra-high-energy cosmic-rays impinging on the Earth's atmosphere with energies far beyond LHC energies. Today’s second-generation of digital radio-detection experiments, as e.g. AERA or LOFAR, are becoming competitive in comparison to already standard techniques e.g. fluorescence light detection.
Thanks to a detailed understanding of the physics of the radio emission in extensive air showers, simulations of the radio signal are already successfully tested and applied in the reconstruction of cosmic rays.
However the limits of the computational power resources are easily reached when it comes to computing electric fields at the numerous positions requested by large or dense antenna arrays. In the case of mountainous areas as e.g. for the GRAND array, where 3D shower simulations are necessary, the problem arises with even stronger acuity.
Therefore we developed a full parametrisation of the emitted radio signal on the basis of generic shower simulations which will reduce the simulation time by orders of magnitudes. In this talk we will present this concept after a short introduction to the concept of the radio detection of air-shower induced by cosmic rays.
\end{abstract}



\section{Introduction}
Thanks to the development in the digital signal processing, the radio detection of air showers which are induced by cosmic rays experienced a renaissance in the last decades~\cite{TimsReview}. Especially, the results of AERA on the energy reconstruction of the primary particle~\cite{AERAEnergy} or LOFAR on the measurement of the mass composition~\cite{LOFARMass} show that radio detection is nowadays competitive to standard detection techniques as e.g. fluorescence light.
These successes are based on the huge progress in the understanding and modeling of the radio emission mechanisms of air showers. 
There are several air-shower simulation programs  on the market. They include modules for the calculation of the corresponding radio emission which differ in their complexity, but all of them take effects on the radio signal, e.g. due to refractive index, into account. In the last years their results started to agree with each other as well as they are consistent with radio-signal measurements under laboratory conditions~\cite{SLAC} or by air-shower arrays, e.g. see~\cite{LOPES}.\\
\\
Nevertheless, running simulations for very high-energetic cosmic rays and arrays consisting of hundreds of antennas, e.g. in case of very large and non-flat arrays, is very CPU-time requesting so that one easily reaches the limitations of the computational resources. 
In the context of GRAND~\cite{GRAND}, a sensitivity study shall be performed to find the locations with an enhanced detection rate of neutrino induced showers, so-called hot spots. This study will consists of running 1 Mio. antennas distributed over 1 Mio. $\mbox{km}^2$. Therefore, running massive simulations is not an option.\\
\\
Within this project, we develop and test a parametrisation of the the radio signal, called radio morphing, to derive the expected electric field emitted from any air shower which is detected at any antenna position from one simulated generic shower. This will be performed by a simple scaling of the electric field amplitude and an interpolation of the pulse shape.

\section{Air shower and radio emission}
\begin{figure}[t]
 \centering
 \includegraphics[width=0.3\textwidth,clip]{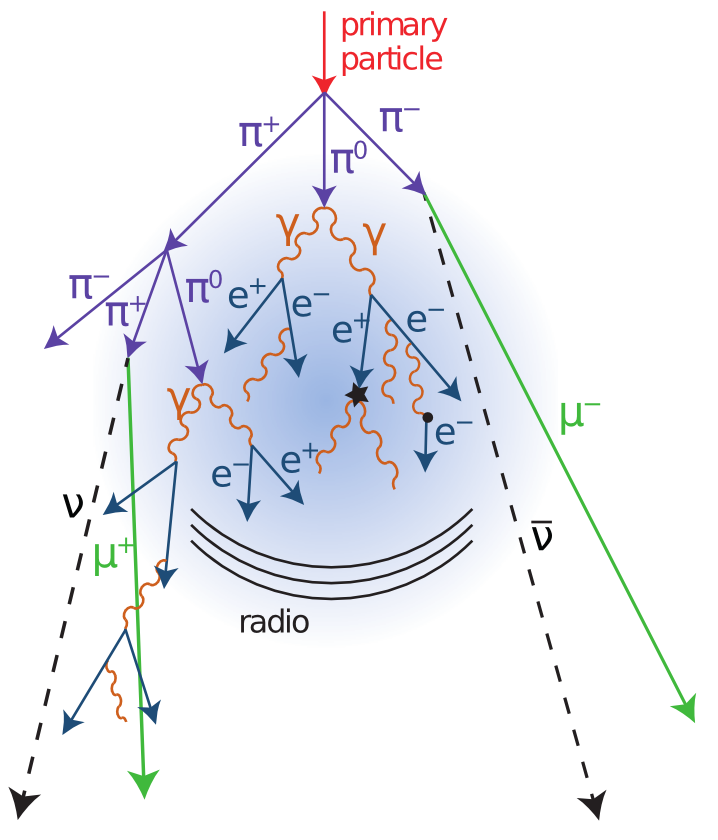}
 \includegraphics[width=0.6\textwidth,clip]{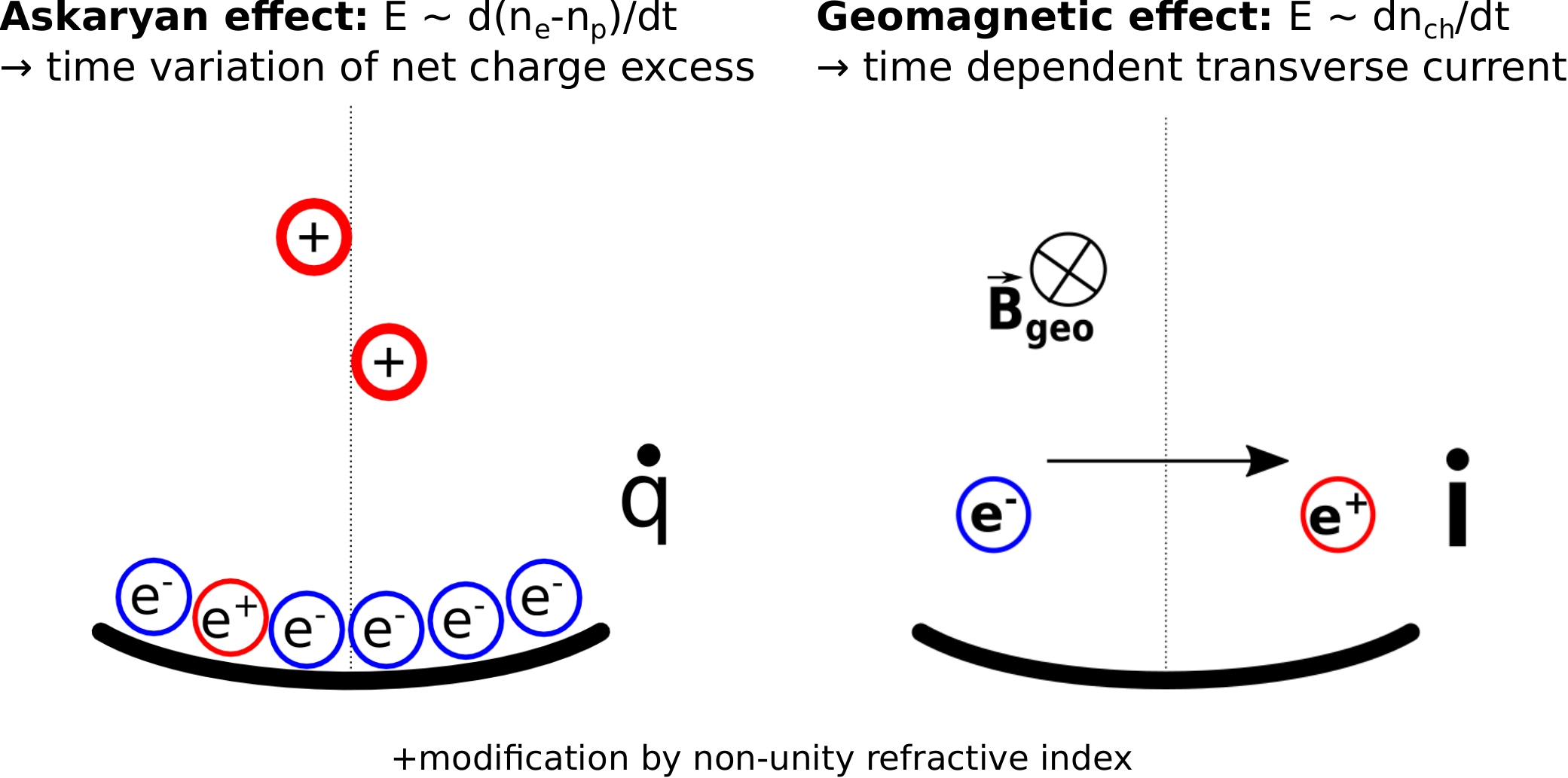}
  \caption{{\bf Left:} The particle components of an air shower~\cite{FranksReview} (modified). {\bf Right:} The radio-emission mechanisms of an air shower.}
  \label{zilles:fig1}
\end{figure}
%
The cascade of ionized particles and electromagnetic radiation in the atmosphere induced
by a primary cosmic particle is also called an extensive air shower. 
Via a nuclear interaction, a primary particle interacts with a nucleus of the Earth's atmosphere and produces secondary particles. These secondaries can be ordered in three groups: the muonic component consists of muons and neutrinos, the hadronic one of protons, neutrons, nuclear fragments, neutral and charged pions and kaons, and the electromagnetic component of electrons, positrons and photons (see fig.~\ref{zilles:fig1}, left).
The geometry of an air shower can be described by following initial shower parameters: the energy of the primary cosmic particle $E$ with which the number of particles in the shower scales, the angles $\theta$ and $\phi$ which describes the direction where the showers goes to, and the injection height $h$ which represents the vertical height of the first interaction from ground.
An observable which can be linked to the mass of the primary particle initiating the extensive air shower is the atmospheric depth $X_{\mathrm{max}}$ (given in $\mbox{g}/\mbox{cm}^2$) at which an air shower reaches its maximum
particle number in the electromagnetic component. Since the strength of the emitted radio signal scales linearly with the number of electrons and positrons, $X_{\mbox{max}}$ can be seen at first order also as the position from where the maximum radiation comes from. \\
\\
The electromagnetic component of this particle shower creates radio emission while propagating through the Earth's atmosphere or through a dense medium. Pulses of a length of tens of nanoseconds are produced, varying in amplitude between pulses hidden in the Galactic noise and pulses with amplitudes orders of magnitudes above it. 
The signal can be interpreted by two main mechanisms for emission of the signal, and an additional modification by the air's refractive index~\cite{TimsReview} (compare to fig.~\ref{zilles:fig1}, right):\\
\\
The first, the so-called Askaryan effect~\cite{Askaryan1962,Askaryan1965}, can be described as a variation of the net charge excess of the shower in time ($\dot{q}$).
Effectively, the surrounding medium is ionized by the air shower particles passing through the Earth's atmosphere and the ionization electrons are swept into the cascade, whereas the heavier positive ions stay behind. 
In a non-absorptive, dielectric medium the number of electrons in the particle front varies in time and therefore the net charge excess in the cascade which develops then a coherent electromagnetic pulse. \\
The second mechanism and the dominant emission mechanism in an air shower is the geomagnetic effect~\cite{TimsReview}:
The secondary electrons and positrons in the shower are accelerated by a geomagnetic field. 
They are decelerated due to the interactions with air molecules. In total, this leads to a net drift of the electrons and positrons in opposite directions as governed by the Lorentz force $F=q \cdot \vec{v} \times \vec{B}$, with $q$ as the particle charge, $\vec{v}$ as the velocity vector of the shower and $\vec{B}$ as the magnetic field vector. As these also referred as "`transverse currents"' vary in time during the air shower development ($\dot{I}$), they lead to the emission of electromagnetic radiation.

\section{Radio morphing}
\begin{figure}[t]
 \centering
 \includegraphics[width=0.9\textwidth,clip]{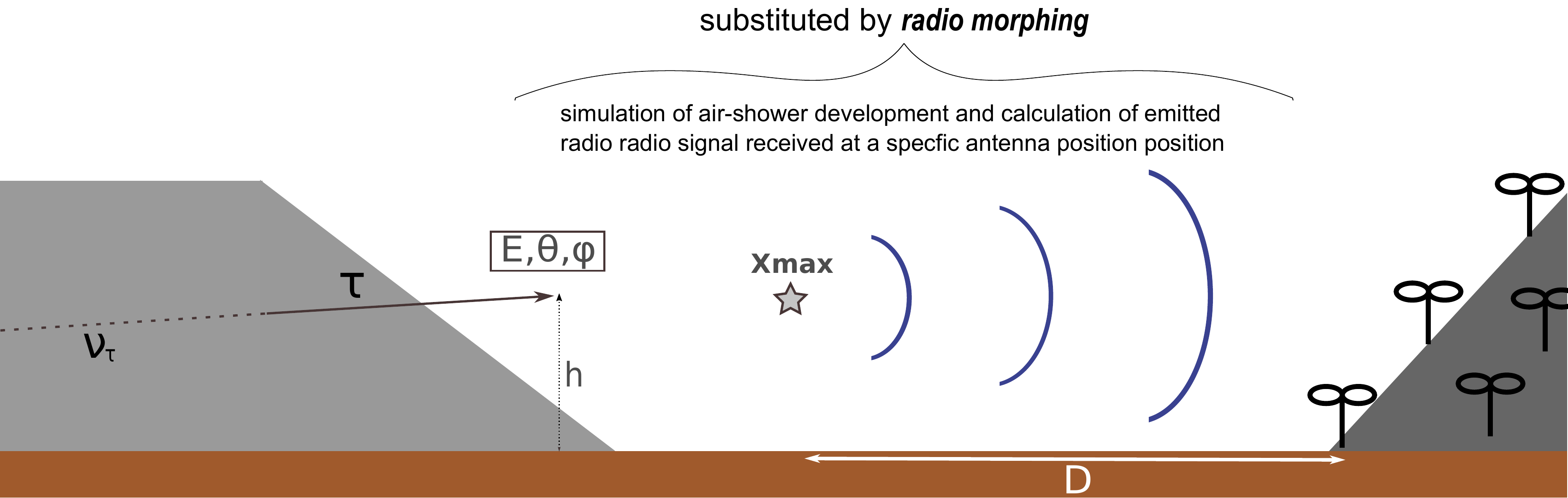}
!
  \caption{The detection chain of a neutrino with GRAND: radio morphing will substitute the actual simulation of each induced air shower.}
  \label{zilles:fig2}
\end{figure}
To be more efficient than simply running massive simulations, we are working on a parametrisation of the radio signal based on the air-shower parameters (see fig.~\ref{zilles:fig2}), so that the expected electric field emitted from any air shower which is detected at any position can be derived from one generic shower simulated with a high antenna density. 
In other words the goal can be defined as following: the electric field $\vec{E}_A(t,x_i)$ at position $x_i$ emitted from the desired target air shower $A$ with given primary parameters $E_A, h_A,\theta_A, \phi_A$ can be derived from the simulated electric field $\vec{E}_B(t,x)$ at position $x$ for the simulated generic air shower $B$ with the primary parameters $E_B, h_B,\theta_B, \phi_B$. This idea is called \textbf{\emph{radio morphing}}. \\
\\
Earlier simulation studies already showed that at fixed distances from the shower maximum, the strength of the radio signal solely depends on $E, \theta, \phi$ and $h$ of the induced air shower as well as that each of these parameter dependencies can be parametrised by simple scaling factors (e.g. $k_E= E_A/E_B$). Furthermore, each of these resulting scaling factors is independent from the others: $k_{AB}=k_E \cdot k_\theta \cdot k_\phi \cdot k_h$.\\  
\\
Based on these results, radio morphing consists of:
\begin{itemize}
 \item \textbf{Producing the generic shower B}\\ 
For the calculation of the radio emission of an air shower, the shower simulation program Aires~\cite{AIRES} and its module for the calculation of the emitted radio signal ZHAireS are used. To achieve the electric field traces at the antenna positions, the antenna positions for the simulation are arranged in planes with different distances $D$ to the shower maximum $X_{\mathrm{max}}$ to account for the dependency of the air-shower development on the air density.

 \item \textbf{Scaling the amplitude of the reference shower $A$ to the desired parameters of the target shower $B$: $\vec{E}_A(x,t) =k_{AB}\cdot \vec{E}_B(x,t)$} \\
For each simulated antenna position, the scaling of the peak amplitudes of the electric field traces is performed accordingly to the desired shower parameters $E_{A}$, $\theta_A$, $\phi_A$ and injection height $h_A$ with respect to the initial shower parameters $E_{B}$, $\theta_B$, $\phi_B$ and $h_B$ by multiplying the factor $k_{AB}=k_E \cdot k_\theta \cdot k_\phi \cdot k_h$. To account for the impact of the refractive index on the actual radio signal and therefore for another signal distribution at the specific distance to the shower maximum, the scaling procedure includes as well a stretching of the antenna positions if needed.
	
\item \textbf{3D interpolation of the electric field trace to the desired antenna position $x_i$: $\vec{E}_A(x,t) \rightarrow \vec{E}_A(x_i,t)$}\\
On the basis of the simulated antenna positions which are organised in planes in specific distances to the shower maximum, the electric field trace of any antenna positions which lies in-between these planes can be derived by an interpolation of the full pulse shape.
The desired electric field traces are calculated in the frequency domain where the spectrum can be represented in polar coordinates: $f(r,\phi)=r\cdot e^{i\phi}$.  The interpolation is performed by applying a linear interpolation of the amplitude $r$ and the phase $\phi$ weighted with the distance of the desired antenna positions to the simulated antenna positions.
\end{itemize}

\section{Summary} 
To perform a simulation study for arrays consisting of hundreds of radio antennas, more efficient simulations are needed. Therefore a parametrisation of the air-shower's radio emission is currently under development and testing. Here, the goal is to compute the expected electric field at any position emitted from any air shower on the basis of a generic air shower as reference which is simulated with a high antenna density. This so-called radio morphing is based on a pulse scaling depending on the initial shower parameters as well as a 3D interpolation of the electric field's pulse shape at any desired antenna position.
Due to the gain of a very large amount of time and costs, this parametrisation will be an extremely powerful method to perform studies on the detection of air showers via their radio signal. Radio morphing will be finally applied in the sensitivity study for GRAND.

\begin{acknowledgements}
This work is supported by the APACHE grant (ANR-16-CE31-0001) of the French Agence Nationale de la Recherche.
\end{acknowledgements}
%

\bibliographystyle{aa}  
\bibliography{zilles} 

\begin{thebibliography}{10}
\expandafter\ifx\csname natexlab\endcsname\relax\def\natexlab#1{#1}\fi

\bibitem[{Aab {et~al.}(2016)}]{AERAEnergy}
Aab, A. {et~al.} 2016, Phys. Rev. Lett., 116, 241101

\bibitem[{Apel {et~al.}(2015)}]{LOPES}
Apel, W. {et~al.} 2015, 75

\bibitem[{Askaryan(1962)}]{Askaryan1962}
Askaryan, G. 1962, Soviet Physics JETP, 14

\bibitem[{Askaryan(1965)}]{Askaryan1965}
Askaryan, G. 1965, Soviet Physics JETP, 21

\bibitem[{Belov {et~al.}(2016)}]{SLAC}
Belov, K. {et~al.} 2016, Phys. Rev. Lett., 116, 141103

\bibitem[{Buitink {et~al.}(2016)}]{LOFARMass}
Buitink, S. {et~al.} 2016, Nature, 531, 70

\bibitem[{Huege(2016)}]{TimsReview}
Huege, T. 2016, Phys. Rept., 620, 1

\bibitem[{Kotera \& Martineau(2017)}]{GRAND}
Kotera, K. \& Martineau, O. 2017, in SF2A 2017 Proceedings

\bibitem[{Schr\"{o}der(2017)}]{FranksReview}
Schr\"{o}der, F.~G. 2017, Prog. Part. Nucl. Phys., 93, 1

\bibitem[{Sciutto(1999)}]{AIRES}
Sciutto, S.~J. 1999, arXiv:astro-ph/9911331

\end{thebibliography}

\end{document}